\documentclass[preprint,showpacs,preprintnumbers,amsmath,amssymb]{revtex4}

\usepackage{graphicx}

\usepackage{amssymb}

\begin{document}

\newcommand{\fu}{$\mathrm{KOs_2O_6}$}
\newcommand{\fua}{$A\mathrm{Os_2O_6}$}
\newcommand{\fuRb}{$\mathrm{RbOs_2O_6}$}
\newcommand{\fuK}{$\mathrm{KOs_2O_6}$}
\newcommand{\fuCs}{$\mathrm{CsOs_2O_6}$}

\newcommand*{\unit}[1]{\,\mathrm{#1}}
\newcommand{\ce}{-\Delta F}  
\newcommand{\sfc}{\gamma}  
\newcommand{\SFC}{\widetilde{\sfc}}  
\newcommand{\Tc}{T_\mathrm{c}}  
\newcommand{\Cp}{C_\mathrm{p}}  
\newcommand{\Hc}{H_\mathrm{c}}  
\newcommand{\svf}{superconducting volume fraction}
\newcommand{\smf}{superconducting mass fraction}
\newcommand{\kB}{k_\mathrm{B}}  
\newcommand{\muB}{\mu_\mathrm{B}}  




\title{Mass enhancement, correlations, and strong coupling superconductivity in the $\beta$-pyrochlore KOs$_2$O$_6$}


\author{M.~Br\"uhwiler}
 \email{markus.bruehwiler@phys.ethz.ch}
 \affiliation{Laboratory for Solid State Physics, ETH Z\"urich, 8093 Z\"urich, Switzerland.}
\author{S.M.~Kazakov}%
 \affiliation{Laboratory for Solid State Physics, ETH Z\"urich, 8093 Z\"urich, Switzerland.}
\author{J.~Karpinski}%
 \affiliation{Laboratory for Solid State Physics, ETH Z\"urich, 8093 Z\"urich, Switzerland.}
\author{B.~Batlogg}%
 \affiliation{Laboratory for Solid State Physics, ETH Z\"urich, 8093 Z\"urich, Switzerland.}

\date{\today}

\begin{abstract}
To assess electron correlation and electron-phonon coupling in the
recently discovered $\beta$-pyrochlores \fuK\ and \fuRb, we have
performed specific heat measurements in magnetic fields up to
$14\unit{T}$. We present data from high quality single crystalline
\fuK, showing that \fuK\ is a strong coupling superconductor with
a coupling parameter $\lambda_\mathrm{ep} \approx 1.0$ to $1.6$
(\fuRb: $\lambda_\mathrm{ep} \approx 1$). The estimated Sommerfeld
coefficient of \fuK, $\sfc=76$ to $110\unit{mJ/(mol\,K^2)}$, is
twice that of \fuRb\ [$\sfc=44\unit{mJ/(mol\,K^2)}$]. Using
strong-coupling corrections, we extract useful thermodynamic
parameters of \fuK. Quantifying $\lambda_\mathrm{ep}$ allows us to
determine the mass enhancement over the calculated band electronic
density of states. A significant contribution in addition to the
electron-phonon term of $\lambda_\mathrm{c}=1.7$ to $4.3$ is
deduced. In an effort to understand the origin of the enhancement
mechanism, we also investigate an unusual energetically low-lying
phonon. There are three phonon modes per \fuRb, suggestive of the
phonon source being the rattling motion of the alkali ion. This
dynamic instability of the alkali ions causes large scattering of
the charge carriers which shows up in an unusual temperature
dependence of the electrical resistivity.
\end{abstract}



\maketitle

\section{Introduction}

Long standing interest in the pyrochlores stems from their
inherent geometrical frustration due to the metal ions forming a
network of corner-sharing tetrahedra. For insulating pyrochlores,
where localized magnetic moments sit on the vertices of these
tetrahedra, competing interactions lead to a wealth of novel
states of fundamental interest. Ground states of infinite
degeneracy are a typical signature of such frustrated
systems.\cite{Moessner2001,Greedan2001,Ramirez1994} While for
insulating systems the intimate connection between spins and the
lattice topology leads to the observed frustration effects, it is
an open question how charge, spin, and lattice degrees of freedom
are coupled in itinerant systems and how they influence each
other.

In this context, the superconductivity recently found in \fuK\ has
been of considerable interest:\cite{YoMuMaHi2004} It has been
thought that a spin liquid ground state resulting from frustration
in these structures might favor superconductivity through singlet
pairing by a resonating valence bond.\cite{Anderson1973} At the
time of its discovery, \fuK\ was the second pyrochlore
superconductor known after
Cd$_2$Re$_2$O$_7$.\cite{HaMuTaSaYaHi2001,SaYoOhKaKaWaMaHaOn2001,JiHeMcAlDrMa2001}
Soon after, the discovery of superconducting
\fuRb\cite{YoMuMaHi2004a,HiYoYaMuMaMu2005} and
\fuCs\cite{YoMuHi2004} followed. Cd$_2$Re$_2$O$_7$ crystallizes in
the so-called $\alpha$-pyrochlore structure, the structure adopted
by numerous insulating pyrochlores, and \fua\ (where \textit{A} =
Cs, Rb, or K) has the $\beta$-pyrochlore structure. While the
\fua\ compounds could be expected to be very similar in their
physical properties, there is in fact quite a variation among
them.

For example, $\Tc$ decreases from $9.5\unit{K}$ for \fuK\ to
$6.4\unit{K}$ for \fuRb\ and to $3.3\unit{K}$ for \fuCs. The
nature of the superconducting state is still under discussion:
while \fuRb\ appears to be a fully gapped $s$-wave superconductor
with a critical temperature $\Tc =
6.4\unit{K}$,\cite{BrKaZhKaBa2004,KhEsKaKaZhBrGaDiShMaMaBaKe2004,MaGaPeHiWeOtKaKa2004,BrKaKaBa2005,MaMiHiBrBa2006}
$\mu$-SR measurements point towards an anisotropic or multigapped
superconducting state in \fuK.\cite{KoHiOhSaKaYoMuHi2005} This
question has not been settled conclusively. It has been found that
\fuK\ single crystals show Bragg peaks that violate the
F\,d\,$\bar{3}$\,m symmetry and the structure has been identified
as non-centrosymmetric F\,$\bar{4}$\,3\,m.\cite{ScKaRoZhKa2006}
Furthermore, it remains to be determined to what degree the
quasiparticle mass is enhanced by interactions other than those
with ordinary phonons, e.g.~to what extent electron-electron
correlations play a role in these compounds. In order to address
these questions a detailed characterization of these compounds is
required.

In this paper, we present field and temperature dependent
thermodynamic and transport measurements on \fuRb\ and \fuK. The
data provide evidence for a significant additional electronic mass
enhancement beyond the contribution from the coupling to phonons
and evidence for low-lying vibrational modes. We discuss
similarities and differences between the two compounds and compile
an extended list of the basic thermodynamic parameters.

\section{Experimental}

\fuK\ single crystals have been grown by an encapsulation
technique in an evacuated quartz tube. In this method, a
stoichiometric amount of Os metal (Alfa Aesar, $99.9\unit{\%}$)
and KO$_2$ (Alfa Aesar, $95\unit{\%}$) is thoroughly mixed in an
argon filled dry box and pressed into a pellet. The pellet with a
mass of $0.2$ to $0.25 \unit{g}$ is put into a quartz tube and
silver oxide Ag$_2$O (Aldrich, $99\unit{\%}$) is added to create
an appropriate oxygen partial pressure during the synthesis. The
evacuated and sealed tube is then inserted into a furnace
preheated to $600\unit{^{\circ}C}$, kept there for $1\unit{h}$,
cooled to $400\unit{^{\circ}C}$ at a rate of
$5\unit{^{\circ}C/h}$, and then cooled to room temperature at a
rate of $150\unit{^{\circ}C/h}$. The resulting black single
crystals of size up to $0.3\unit{mm}$ in diameter have grown on
the wall of the quartz tube and on the surface of the precursor
pellet. The crystals have different shapes characteristic for the
cubic structure such as octahedra or cubes. The lattice parameter
refined from x-ray single crystal measurements is found to be $a =
10.0968(8)\unit{\AA}$. No signature of impurity phases like
OsO$_2$ or KOsO$_4$ was found in heat capacity and magnetization
measurements.

For the specific heat measurements, we have selected about five
dozen of the most regularly shaped single crystals out of a single
badge. The total mass of these crystals amounts to $2.12\unit{mg}$
(see inset to Fig.~\ref{fig:CvsHandCrystals}), which is enough to
measure specific heat with a relaxation method. Special attention
must be paid, however, to measure the addendum contribution as it
makes up a significant fraction of the total mass. The specific
heat has been measured in a physical properties measurement
apparatus using an adiabatic relaxation technique (Quantum Design,
PPMS). Resistivity has been measured by a standard four terminal
approach also using the PPMS and the dc magnetization has been
measured in a SQUID magnetometer (Quantum Design, MPMS).
Measurements on \fuRb\ have been performed on a sample previously
reported upon in Ref.~\cite{BrKaKaBa2005}. The synthesis procedure
of these polycrystalline \fuRb\ samples is described
elsewhere.\cite{YoMuMaHi2004,KaZhBrBaKa2004}

\section{Results}

In order to determine the mass enhancement in the normal state and
the magnitude of the electron-phonon coupling, it is necessary to
measure the Sommerfeld coefficient of the superconductor by
driving it into the normal state. Usually this is achieved by
applying a magnetic field exceeding the upper critical field
$H_\mathrm{c2}$. Owing to a high $H_\mathrm{c2}(0 \unit{K})$ of
\fuK\ which is unaccessible by our $14\unit{T}$ magnet, it is not
possible to tune \fuK\ into its normal state below about
$6.2\unit{K}$. We therefore resort to an extrapolation method to
estimate the Sommerfeld coefficient: first we measure the phase
boundary line $H_\mathrm{c2}(T)$ to as low a temperature and high
field as possible. We then extrapolate $H_\mathrm{c2}$ to
$0\unit{K}$ and estimate the Sommerfeld coefficient based on the
measured $\Cp/T$ versus magnetic field curve at a low temperature.
The details of this procedure are given in the following.

\subsection{Upper critical field $H_\mathrm{c2}$}

\begin{figure}
  \begin{center}
  \includegraphics{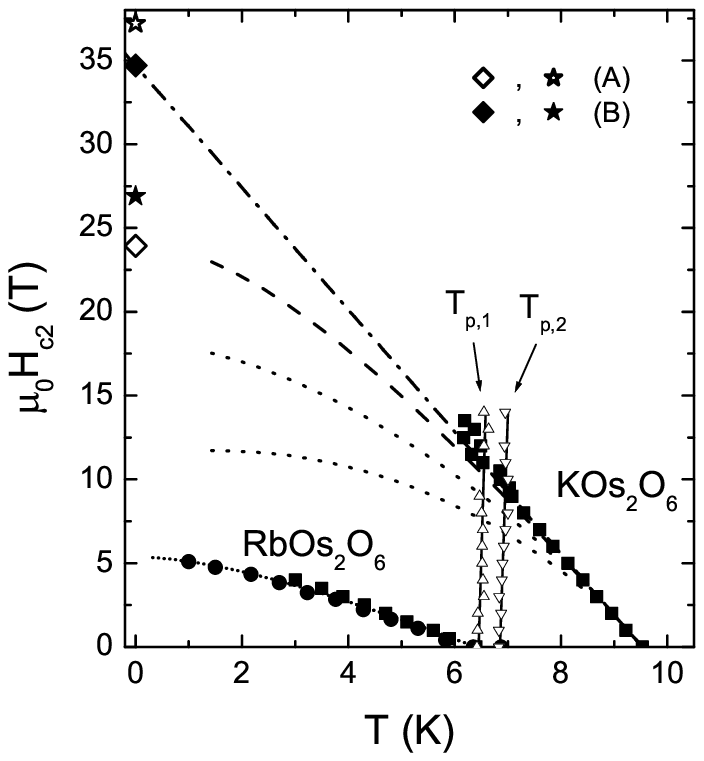}
  \caption{\label{fig:Hc2} The upper critical field
$H_\mathrm{c2}(T)$ of \fuK\ and \fuRb. The squares
($\blacksquare$) are data points extracted from $\Cp/T$ versus $T$
data, while the circles ($\bullet$) are extracted from $\Cp/T$
versus $H$ data. For \fuK, two methods to extrapolate to
$0\unit{K}$ are used: (A) The dashed line is from the WHH formula
in the orbital limit ($\lambda_\mathrm{SO}=\infty$), providing a
good description of the data up to about $9\unit{T}$, while a
description with smaller spin-orbit scattering strength (dotted
lines) deviates markedly. (B) The dash-dotted line is a linear fit
to the data. The stars mark the Pauli-limiting field $\mu_0
H_\mathrm{P}$ for \fuK. $T_\mathrm{p,1}$ and $T_\mathrm{p,2}$ mark
two additional peaks observed in the specific heat
(Fig.~\ref{fig:DeltaCatTc}) that might be associated with K
ordering.}
  \end{center}
\end{figure}

We extract the upper critical field $H_\mathrm{c2}$ from $\Cp(T)$
data for \fuK\ and from both $\Cp(T)$ and $\Cp(H)$ data for \fuRb.
For both methods, we use a construction of equal area to determine
the critical field. The two methods result in almost identical
$H_\mathrm{c2}$ as can be seen in Fig.~\ref{fig:Hc2}. In our
experience, the $H_\mathrm{c2}$ slope at $\Tc$ for both \fuRb\ and
\fuK\ are rather stable quantities varying little from sample to
sample. We therefore have a high level of confidence that the
observed slope is indeed intrinsic to the material and not mainly
limited by a mean free path caused by impurity scattering.


The initial slope of the critical boundary at $T_\mathrm{c}$ for
\fuK\ is $-3.6\unit{T/K}$, in agreement with the revised value in
Ref.~\cite{HiYoYaMuMu2005,HiYoYaMuMaMu2005}. The full
$H_\mathrm{c2}(T)$ dependence cannot be mapped out with the
$14\unit{T}$ magnet, so we need to extrapolate to lower
temperatures and we proceed in two different ways: first, we
notice that $H_\mathrm{c2}(T)$ in \fuRb\ is rather well described
by the standard WHH expression, with an initial slope of
$-1.2\unit{T/K}$ and $\mu_0 H_\mathrm{c2}(0\unit{K}) = 6\unit{T}$
(Fig.~\ref{fig:Hc2}).\cite{BrKaZhKaBa2004} If we were to assume
the same functional form to hold also for \fuK, $\mu_0
H_\mathrm{c2}(0\unit{K}) = 24\unit{T}$ is estimated. Considering
that a typical energy of low-lying phonons in \fuK\ is around
$30\unit{K}$, $H_\mathrm{c2}$ might be underestimated in this way.
Alternatively, one can extrapolate linearly to $T=0\unit{K}$ and
would estimate $\mu_0 H_\mathrm{c2}(0\unit{K}) = 35\unit{T}$.
Pursuing this second extrapolation further is motivated by our
observation that the last data points in our data lie closer to a
straight line extrapolation of the lower-field data, while the WHH
curve starts deviating from the data in this region. In the rest
of this paper, we thus consider both extrapolations, denoting with
(A) the WHH version, and with (B) the linear method. Numerical
values for (B) will be enclosed in braces \{\}.

It is worth noting that the \fuRb\ data are fitted best by the WHH
formula in the orbital limit ($\lambda_\mathrm{SO} = \infty$).
This is in agreement with a large $\lambda_\mathrm{SO}$ expected
for the heavy Os ion. Therefore, pair-breaking effects of the
magnetic field on the spin degrees of freedom are negligible due
to the randomizing effect of the spin-orbit scattering on the
phase of the superconducting electrons. An upper limit of the
critical field in the case where spin-orbit scattering could be
ignored, is given by the Pauli-limiting field: for \fuK\ it is
$\mu_0 H_\mathrm{P} = \Delta(0\unit{K})/(\sqrt{2}\muB)
\sqrt{1+\lambda_\mathrm{ep}} \approx 37\unit{T}$ \{$27\unit{T}$\},
while for \fuRb\ it is about $18\unit{T}$. Here we have used the
values for $\Delta(0\unit{K})$ and $\lambda_\mathrm{ep}$
determined from our measurements as described later and listed in
Table \ref{tab:TD_params}. For \fuRb, this field is well above the
measured upper critical field, while for \fuK\ the comparison
depends on the extrapolation of $H_\mathrm{c2}$. Support for the
fact that $H_\mathrm{c2}$ is determined by the orbital motion of
the electrons and not by their spin, is provided by the ratio of
the slopes $-\mathrm{d}\mu_0
H_\mathrm{c2}/\mathrm{d}T\vert_{T_\mathrm{c}}$ for the two
materials \fuRb\ and \fuK. The slopes are proportional to $1/(\tau
v_\mathrm{F}^2)$, where $\tau$ is the electronic scattering time
and $v_\mathrm{F}$ the Fermi velocity. For the Fermi velocities we
use $2.587 \cdot 10^7\unit{cm/s}$ for \fuRb\ and $2.671 \cdot
10^7\unit{cm/s}$ for \fuK.\cite{Sa2005} The ratio of the Fermi
velocities squared, taking into account the renormalization of the
Fermi velocity and assuming the same $\tau$ for both materials, is
$2.4$ \{$4.7$\}. This is indeed close to the ratio of
$-\mathrm{d}\mu_0 H_\mathrm{c2}/\mathrm{d}T\vert_{T_\mathrm{c}}$
for \fuRb\ and \fuK: $3.6/1.2 = 3$, suggesting that the
pair-breaking is caused conventionally by the effect of the
magnetic field on the orbital magnetism of the electrons. The
difference in magnitude of the upper critical fields is therefore
in line with the different mass enhancements.

\subsection{Heat capacity in a magnetic field}

We have measured the field dependence of $\Cp/T$ of \fuK\ at
$0.46\unit{K}$, shown in Fig.~\ref{fig:CvsHandCrystals}. The
addendum is $\approx 10\unit{\%}$ of the total heat capacity at
$\mu_0 H = 14\unit{T}$, $\approx 20\unit{\%}$ at $\mu_0 H =
7\unit{T}$, and $\approx 50\unit{\%}$ at $\mu_0 H = 3\unit{T}$.
Below $\mu_0 H = 1.7\unit{T}$, the addendum makes up more than
$90\unit{\%}$ of the total heat capacity, and even though both
data sets are very smooth, this leads to relatively larger scatter
after taking the difference of the two data sets. For this reason,
data below $1.7\unit{T}$ have been omitted in the graph and in the
fitting procedure.

We parameterize the measured curve by $\Cp / T = a
(H/\mathrm{Oe})^b + c$. A least squares fit down to $1.75
\unit{T}$ using this form results in
$a=(5.86\pm1.3)\cdot10^{-4}\unit{mJ/(mol\,K^2)}$,
$b=0.9512\pm0.02$, and $c=-0.35\pm0.4 \unit{mJ/(mol\,K^2)}$, where
$H$ is given in Oersted. An extrapolation of the $\Cp(H)$ curve at
$0.46\unit{K}$ to the upper critical field $H_\mathrm{c2}$ of
$24\unit{T}$ \{$35\unit{T}$\} results in a rather high Sommerfeld
coefficient for \fuK\ of $\sfc=76\unit{mJ/(mol\,K^2)}$
\{$110\unit{mJ/(mol\,K^2)}$\}.

The fact that we have measured a vanishingly small $\Cp/T$ at
$H=0\unit{Oe}$ and $T=0.46\unit{K}$ is of significance for the
interpretation of the electronic and vibrational excitation
spectrum. First, it indicates that the superconducting state
affects the entire electronic system that gives rise to the large
$\sfc$ value [$76$ to $110\unit{mJ/(mol\,K^2)}$]. Secondly, the
vibrational excitation spectrum (at $0.46\unit{K}$) is not
characterized by a glass-like spectrum ($\Cp \propto T$) as one
might expect from a broad distribution of $2$-level configurations
possibly associated with the particular lattice potential
experienced by the K ions.\cite{KuJePi2004}

In addition to estimating the Sommerfeld coefficient, measurements
of the magnetic field dependent heat capacity can be helpful to
reveal nodes in the gap function. It is generally thought that a
square root dependence $\Cp/T \propto H^\alpha$ with $\alpha =
1/2$ indicates a gap with nodes, while a linear dependence
$\alpha=1$ indicates a fully gapped state.\cite{Volovik1993} This
is because the electronic excitations in the vortex state of
conventional superconductors are predominantly low-energy
excitations localized in the vortex core. On the other hand, the
density of states for superconductors with lines of gap nodes
results mostly from delocalized states outside the vortex which
are located in the vicinity of the gap nodes in momentum space.

We expect that the exponent in the above fit to $\Cp$ versus $H$
tends even closer toward $1$ as $T\to0\unit{K}$, since the
exponent decreases when measuring at increasingly higher
temperatures (data not shown). The field-dependence of the heat
capacity of \fuK\ thus appears to be in line with a fully gapped
state. However, since the measurements do not extend up to
$H_\mathrm{c2}$, further experimental evidence is needed to settle
this point conclusively.

\subsection{Density of states and mass enhancement}

\begin{figure}
  \begin{center}
  \includegraphics{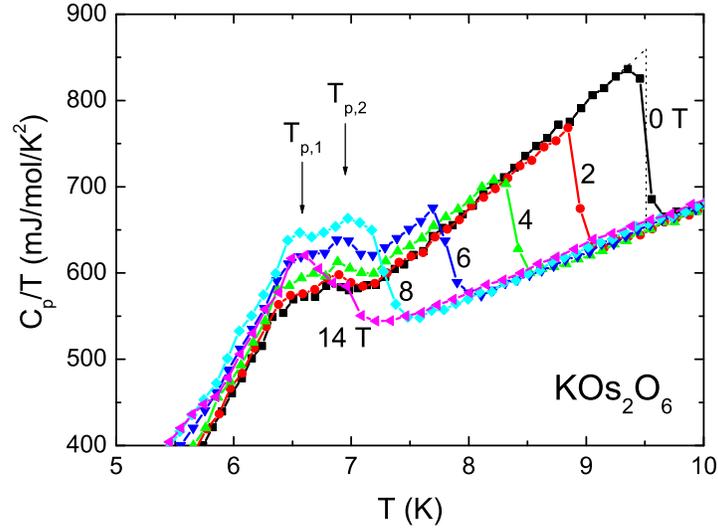}
  \caption{\label{fig:DeltaCatTc} (Color online) Heat capacity versus temperature of \fuK\ for various
  magnetic fields. The specific heat jump at $\Tc$, $\Delta
\Cp\vert_{T_\mathrm{c}}/\Tc$ is $204\unit{mJ/(mol\,K^2)}$, from
which the normalized specific heat jump $\Delta
\Cp\vert_{T_\mathrm{c}}/(\sfc \Tc) = 2.7$ \{$1.9$\} is extracted.
It is significantly larger than that for the weak-coupling case
and corresponds to an electron-phonon coupling constant
$\lambda_\mathrm{ep} \approx 1.0$ \{$1.6$\}, i.e.~\fuK\ is a
superconductor in the strong-coupling regime. There are two
transitions at $T_\mathrm{p,1} \approx 6.5\unit{K}$ and
  $T_\mathrm{p,2} \approx 7\unit{K}$ that might be associated with K ordering.}
  \end{center}
\end{figure}

The specific heat for \fuK\ in various fields is shown in
Fig.~\ref{fig:DeltaCatTc}. In addition to the anomaly indicating
the transition into the superconducting state, there are two
additional peaks at $T_\mathrm{p,1}$ ($\approx 6.5\unit{K}$) and
$T_\mathrm{p,2}$ ($\approx 7\unit{K}$). The origin of these peaks
is unclear at this point. They are likely to be associated with
the dynamics of the potassium ions in this compound. While we
measure two peaks for our sample, a single peak has been reported
in Ref.~\cite{HiYoYaMuMu2005,HiYoYaMuMaMu2005} at a slightly
higher temperature (about $7.5\unit{K}$) - the transition seems to
be very sensitive to microscopic details.

From the specific heat jump at $\Tc$, $\Delta
\Cp\vert_{T_\mathrm{c}}/\Tc = 204 \unit{mJ/(mol\,K^2)}$, the
normalized specific heat jump $\Delta
\Cp\vert_{T_\mathrm{c}}/(\sfc \Tc) = 2.7$ \{$1.9$\} is extracted.
It is significantly larger than in the weak-coupling limit and
corresponds to an electron-phonon coupling constant
$\lambda_\mathrm{ep} = 2 \int_0^\infty \alpha^2 F(\omega)/\omega
\, \mathrm{d}\omega \approx 1.6$ \{$1.0$\},\cite{MaCoCa1987}
i.e.~\fu\ is a superconductor in the strong-coupling regime. Here,
$\alpha^2 F(\omega)$ is the spectral density of the
electron-phonon coupling function. Using the calculated band
Sommerfeld coefficient $\sfc_\mathrm{b} = 9.8$ to
$11.36\unit{mJ/(mol\,K^2)}$ of \fuK\ from
Refs.~\cite{KuJePi2004,SaFr2005}, this result indicates a
significant enhancement of the electronic specific heat of
$(1+\lambda_\mathrm{ep})(1+\lambda_\mathrm{c}) =
[76\unit{mJ/(mol\,K^2)}]/[10.6\unit{mJ/(mol\,K^2)}] \approx 7.2$
\{$10.4$\}, i.e.~about double the enhancement found in
Sr$_2$RuO$_4$ of $3.8$ to $4$.\cite{Singh1995,Oguchi1995} This
quantification of $\lambda_\mathrm{ep}$ then leaves a significant
additional enhancement $\lambda_\mathrm{c} \approx 1.7$ \{$4.3$\}
ascribed to electron-electron correlations.

\begin{figure}
\begin{center}
  \includegraphics[width=0.6\textwidth]{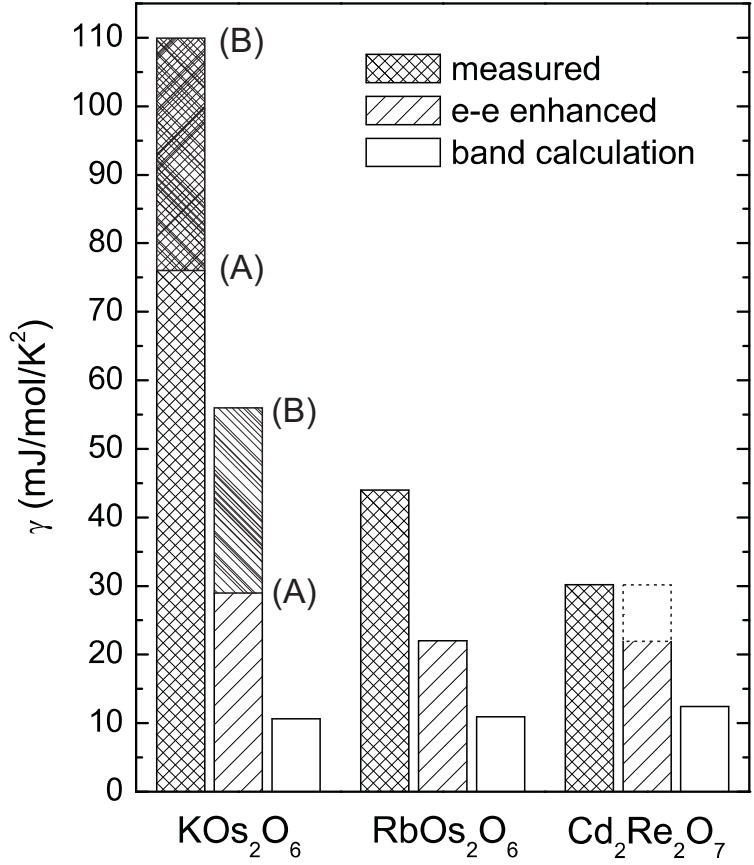}
  \caption{\label{fig:DOSGraph} Comparison of the Sommerfeld
coefficients for \fuRb, \fuK, and the weak-coupling
$\alpha$-pyrochlore superconductor Cd$_2$Re$_2$O$_7$. Shown are
the calculated bare coefficients $\sfc_\mathrm{b}$ in white, the
electron-electron enhanced coefficients
$(1+\lambda_\mathrm{c})\sfc_\mathrm{b}$ in hatched, and the
measured coefficients
$(1+\lambda_\mathrm{ep})(1+\lambda_\mathrm{c})\sfc_\mathrm{b}$ in
cross hatched. The bare band value from DFT calculations in the
LDA is rather similar for all three pyrochlores. The
electron-phonon coupling $\lambda_\mathrm{ep}$ increases from $<
0.4$ (weak coupling) for Cd$_2$Re$_2$O$_7$ to $1$ for \fuRb\ to
$1.0$ to $1.6$ for \fuK. The two sets of values (A) and (B) for
\fuK\ correspond to the two choices of extrapolating
$H_\mathrm{c2}(T)$ in Fig.~\ref{fig:Hc2}. }
\end{center}
\end{figure}

Figure \ref{fig:DOSGraph} illustrates the density of states for
\fuK\ and \fuRb\ and the weak-coupling $\alpha$-pyrochlore
superconductor Cd$_2$Re$_2$O$_7$ for comparison. The depicted band
density of states for \fua\ is the arithmetic mean of the two
calculated values in Refs.~\cite{KuJePi2004,SaFr2005}, and the
values for Cd$_2$Re$_2$O$_7$ are taken from
Refs.~\cite{SiBlSchSo2002,HaMuTaSaYaHi2001}. It increases very
slightly on going from $A$=K to $A$=Rb to Cd$_2$Re$_2$O$_7$, but
is rather similar overall. The electronic structure in the LDA as
a whole is in fact very similar for all \fua. The electron-phonon
enhancement, on the other hand varies significantly:
$\lambda_\mathrm{ep}$ is less than about $0.4$ for
Cd$_2$Re$_2$O$_7$, but falls in the intermediate to
strong-coupling range for the osmates ($1$ to $1.6$), indicating
that there is a significant difference in the coupling functions
leading to the superconducting state. The difference in the
additional enhancement $\lambda_\mathrm{c}$ is pronounced: it
increases more than twofold from \fuRb\ to \fuK, in the same
direction as $\lambda_\mathrm{ep}$.

\begin{figure}
\begin{center}
  \includegraphics[width=0.7\textwidth]{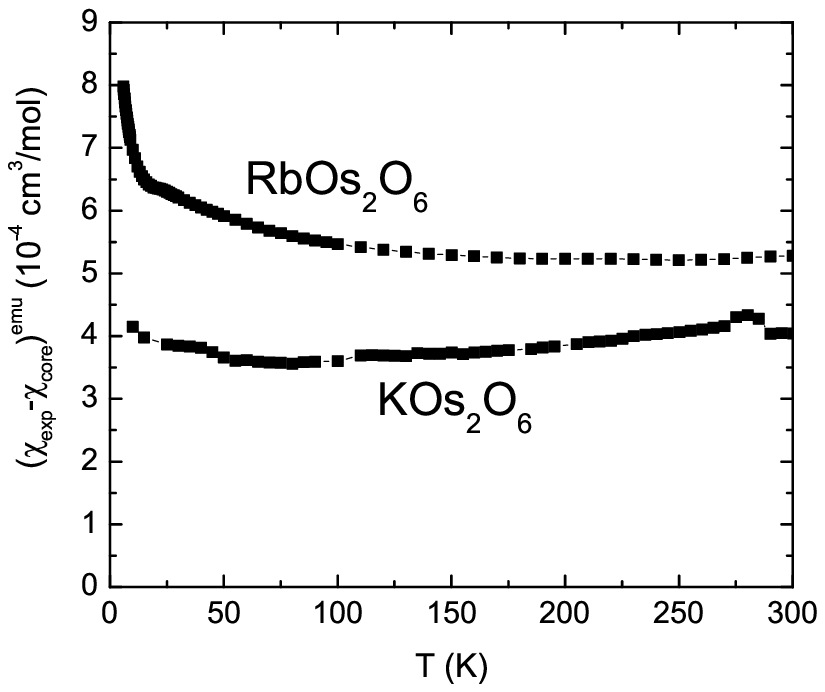}
  \caption{\label{fig:chi_KOs2O6_n_RbOs2O6} Magnetic susceptibility
of \fuK\ and \fuRb\ measured at $1\unit{T}$. The data is corrected
for the core diamagnetism ($+1.22\cdot10^{-4}\unit{cm^3/mol}$ for
\fuRb, $+1.15\cdot10^{-4}\unit{cm^3/mol}$ for \fuK), resulting in
susceptibilities taken at $150\unit{K}$ of $\chi^\mathrm{emu}
\approx 3.7 \cdot 10^{-4} \unit{cm^3/mol}$ for \fuK\ and
$\chi^\mathrm{emu} \approx 5.3 \cdot 10^{-4} \unit{cm^3/mol}$ for
\fuRb.}
\end{center}
\end{figure}

To gain further insight into the mechanisms responsible for the
density of states enhancement, it is helpful to consider the
magnetic susceptibility (Fig.~\ref{fig:chi_KOs2O6_n_RbOs2O6}).
Since samples of \fuRb\ also contain some OsO$_2$, the intrinsic
magnetic susceptibility is calculated according to $\chi_1 =
\eta_m^{-1} \chi - (\eta_m^{-1}-1) \chi_2$, where $\chi_1$ is the
magnetic susceptibility of \fuRb, $\chi_2$ that of OsO$_2$, and
$\chi$ is the measured susceptibility of the mixed system
(c.f.~Sec.~\ref{sec:LowEEinsteinPhonon}). The susceptibilities in
this formula are given per mass. The correction due to OsO$_2$ is
very small, since the susceptibility per Os is very similar in
both systems. Even after taking into account the correction for
OsO$_2$, the susceptibility of \fuRb\ still shows a slight
increase at low temperatures and also a shoulder around
$25\unit{K}$. This is probably due to minute amounts of magnetic
RbOsO$_4$. We note that a similar $T$ dependence as for the
susceptibility has been observed for the Knight shift in both
\fuRb\ and \fuK.\cite{ArKiKoTaYoMuHi2004}

We correct for the diamagnetism of the core using $\chi =
\chi_\mathrm{exp} - \chi_\mathrm{core}$ with $\chi_\mathrm{core}$
= $-12$, $-20$, $-13$, and $-18 \unit{\mu cm^3/mol}$ for O$^{2-}$,
Rb$^{+}$, K$^{+}$, and Os$^{6+}$ ions respectively. Landau
diamagnetism can be neglected, due to the enhancement of the
electron mass. For the experimental susceptibility, we use the
value at $150\unit{K}$, resulting in a susceptibility for \fuK\ of
$\chi^\mathrm{emu} \approx 3.7 \cdot 10^{-4} \unit{cm^3/mol}$. The
Sommerfeld-Wilson ratio evaluates to $R_\mathrm{W} =
0.93\unit{G^2\,cm^3/erg}$ \{$0.65\unit{G^2\,cm^3/erg}$\} where we
have used the renormalized Sommerfeld coefficient
$\sfc/(1+\lambda_\mathrm{ep})$ since the Pauli magnetic
susceptibility is not affected by electron-phonon interactions.
For \fuRb\ follows $\chi^\mathrm{emu} \approx 5.3 \cdot 10^{-4}
\unit{cm^3/mol}$ and $R_\mathrm{W} = 1.75\unit{G^2\,cm^3/erg}$. In
view of a calculated Stoner enhancement of the magnetic
susceptibility of roughly $2$ for all \fua\
(Refs.~\cite{KuJePi2004,SaFr2005}), the Wilson ratio for \fuRb\
closely matches the expected result. On the other hand,
$R_\mathrm{W}$ for \fuK\ is clearly smaller than what would be
expected. In this simple estimate, we have neglected possible
orbital van Vleck terms that may also contribute to the
paramagnetic susceptibility.

\subsection{Superconducting Properties}

It is instructive to calculate the various thermodynamic
parameters for \fuK\ and \fuRb. We get a strong-coupling parameter
$x := k_\mathrm{B}\Tc/(\hbar \omega_\mathrm{ln}) = 0.13$
\{$0.06$\} by applying the approximate semiphenomenological form
of the strong-coupling correction to the weak coupling BCS ratio
which holds for many superconductors $\Delta
\Cp\vert_{T_\mathrm{c}} / ( \sfc \Tc ) = 1.43 \left [ 1 - 53 x^2
\ln \left ( 3x \right ) \right ]$.\cite{MaCa1986} Here $\hbar
\omega_\mathrm{ln}$ is the Allen-Dynes expression for the average
phonon energy. With this strong-coupling parameter we further get
a normalized energy gap of $2\Delta(0\unit{K}) / (k_\mathrm{B}\Tc)
= 3.53 \left [ 1 - 12.5 x^2 \ln \left ( 2x \right ) \right ] =
4.57$ \{$3.83$\}. Since we know from \fuRb\ that another
strong-coupling correction, $1/(8\pi) \cdot \sfc\Tc^2/(\ce) =
0.168 \left [ 1 + 12.2 x^2 \ln \left ( 3x \right ) \right ]$,
holds quite well,\cite{BrKaKaBa2005} it is reasonable to assume
that it also holds for \fuK. Using the experimental $\Tc$, $\sfc$,
and the strong-coupling parameter from above, we can estimate the
condensation energy of \fuK: $\ce = 2050\unit{mJ/mol}$
\{$2533\unit{mJ/mol}$\}, corresponding to a thermodynamic critical
field of $\Hc = 2579\unit{Oe}$ \{$2867\unit{Oe}$\}. In this
conversion we have used the calculated mass density of \fuK\ using
the lattice constant from x-ray, $\rho=6.653\unit{g/cm^3}$. The
ratio $1/(8\pi) \cdot (\sfc\Tc^2)/(\ce)$ itself evaluates to
$0.135$ \{$0.157$\}.

\begin{table}
\caption{\label{tab:TD_params} Thermodynamic parameters of the
superconductors \fuK\ and \fuRb. }
\begin{tabular}{lrr}
Parameter & \fuK\ & \fuRb\ \\
\hline
$\Tc$ & $9.5\unit{K}$ & $6.4\unit{K}$\\
$\xi(0\unit{K})$ & $37\unit{\AA}$ \{$31\unit{\AA}$\} & $74\unit{\AA}$\\
$\lambda_\mathrm{eff}(0\unit{K})$ & $243\unit{nm}$ \{$265\unit{nm}$\} & $252\unit{nm}$ \\
$\kappa(\Tc)$, $\kappa(0\unit{K})$ & $45$, $66$ \{$45$, $86$\} & $23$, $34$ \\
\hline
$\sfc$ & $76$ \{$110$\} $\unit{mJ/(mol_\mathrm{f.u.}\,K^2)}$ & $44\unit{mJ/(mol_\mathrm{f.u.}\,K^2)}$ \\
$\Delta \Cp\vert_{T_\mathrm{c}}/(\sfc \Tc)$ & 2.7 \{$1.9$\} & 1.9 \\
$\lambda_\mathrm{ep}$ & $1.6$ \{$1.0$\} & $1.0$ \\
$\lambda_\mathrm{c}$ & $1.7$ \{$4.3$\} & $1.0$ \\
$b$ & $0.95\unit{K^{-1}}$ \{$0.77\unit{K^{-1}}$\} & $1.12\unit{K^{-1}}$ \\
$b/\Tc$ & $0.099\unit{K^{-2}}$ \{$0.080\unit{K^{-2}}$\} & $0.175\unit{K^{-2}}$ \\
$b\Tc$ & $9.021$ \{$7.299$\} & $7.168$ \\
$\ce(0\unit{K})$ & $2050$ \{$2533$\} $\unit{mJ/mol_\mathrm{f.u.}}$ & $483\unit{mJ/mol_\mathrm{f.u.}}$ \\
\hline
$\Hc(0\unit{K})$ & $2579\unit{Oe}$ \{$2867\unit{Oe}$\} & $1249\unit{Oe}$ \\
$H_\mathrm{c1}(0\unit{K})$ & $116\unit{Oe}$ \{$105\unit{Oe}$\} & $92\unit{Oe}$ \\
$\mu_0 H_\mathrm{c2}(0\unit{K})$ & $24\unit{T}$ \{$35\unit{T}$\} & $6\unit{T}$ \\
$-\mathrm{d}\Hc/\mathrm{d}T\vert_{T_\mathrm{c}}$ & $575 \unit{Oe/K}$ & $369 \unit{Oe/K}$ \\
$-\mathrm{d}\mu_0 H_\mathrm{c2}/\mathrm{d}T\vert_{T_\mathrm{c}}$ & $3.6 \unit{T/K}$ & $1.2 \unit{T/K}$ \\
$Q \equiv - \frac{2\Tc}{\Hc(0)} \left.\frac{\mathrm{d}\Hc}{\mathrm{d}T}\right\vert_{T_\mathrm{c}} $ & $4.25$ \{$3.82$\} & $3.79$ \\
$k_\mathrm{B}\Tc/(\hbar\omega_\mathrm{ln})$ & $0.13$ \{$0.06$\} & $0.06$ \\
$2\Delta(0\unit{K}) / (k_\mathrm{B}\Tc)$ & $4.57$ \{$3.83$\} & $3.87$\\
$1/(8\pi) \cdot \sfc\Tc^2 / (\ce)$ & $0.13$ \{$0.16$\} & $0.15$\\
\end{tabular}
\end{table}

Once the condensation energy is known, various thermodynamic
quantities can be evaluated: for $b\Tc \equiv (T\Delta
\Cp)\vert_{T_\mathrm{c}}/(\ce)$ we get $9.02$ \{$7.30$\}, giving a
normalized critical field slope $Q = 4.25$ \{$3.82$\}. The
critical field slope itself evaluates to
$-\mathrm{d}\Hc/\mathrm{d}T\vert_{T_\mathrm{c}} = 575
\unit{Oe/K}$. The slope of the upper critical field is about $3$
times steeper than the one from \fuRb: with $-\mathrm{d}\mu_0
H_\mathrm{c2}/\mathrm{d}T\vert_{T_\mathrm{c}} = 3.6 \unit{T/K}$
(Fig.~\ref{fig:Hc2}) we get a Ginzburg-Landau parameter at the
critical temperature of $\kappa(\Tc) = 45$. At $T \to 0\unit{K}$,
we estimate $\kappa(0\unit{K})= 1/\sqrt{2} \cdot H_\mathrm{c2}/\Hc
= 66$ \{$86$\} and thus a penetration depth of $243\unit{nm}$
\{$265\unit{nm}$\}, which compares well with results from $\mu$-SR
experiments ($270\unit{nm}$).\cite{KoHiOhSaKaYoMuHi2005} The
Ginzburg-Landau coherence length amounts to $\xi = 37\unit{\AA}$
\{$31\unit{\AA}$\} and the lower critical field $H_\mathrm{c1} =
\ln \kappa / (\sqrt{2}\kappa)\Hc = 116\unit{Oe}$
\{$105\unit{Oe}$\}. This lower critical field is in agreement with
magnetization measurements (not shown). The Ginzburg-Landau
coherence length is about half that of \fuRb. The parameters are
listed in Table \ref{tab:TD_params}, where they are compared to
the values of \fuRb\, taken from Ref.~\cite{BrKaKaBa2005} for
convenience.

$H_\mathrm{c2}$ for \fuK\ is beyond the $14\unit{T}$ accessible in
our current setup, so that the samples cannot be tuned to the
normal state below about $6.2\unit{K}$. There is thus no reference
measurement for the heat capacity in the normal state and the
superconducting electronic specific heat $C_\mathrm{es}$ cannot be
determined by a simple subtraction of a normal state and a
superconducting state measurement. This renders determining
additional contributions to the heat capacity other than the usual
phononic and electronic terms difficult.

\subsection{Unusual low-energy atomic vibrations}
\label{sec:LowEEinsteinPhonon}

The specific heat of \fuRb\ is analysed using the condensation
energy analysis (CEA) developed in Ref.~\cite{BrKaKaBa2005}. This
is necessary because it is not yet possible to synthesize a fully
phase pure \fuRb\ sample. OsO$_2$ has been identified by x-ray
diffraction analysis as a secondary phase. This shortcoming is
compensated by the fact that by using the CEA it is possible to
extract the intrinsic properties of superconducting \fuRb\ even if
the samples contain some unreacted OsO$_2$ starting material.

\begin{figure}
  \begin{center}
  \includegraphics{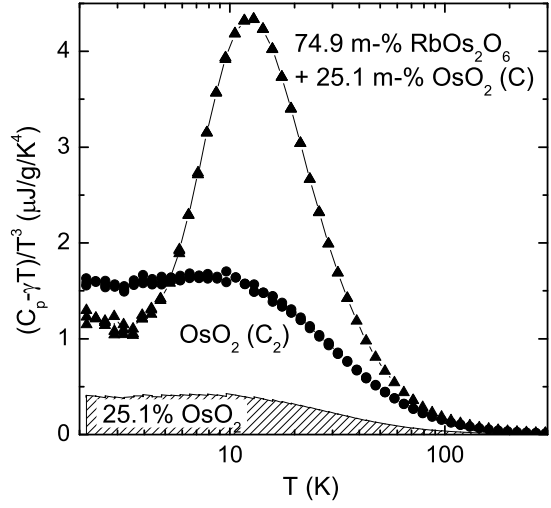}
  \caption{\label{fig:CdivT3_vs_logT_Rb_and_OsO2} $(\Cp - \sfc
T)/T^3$ versus temperature on a logarithmic scale of a \fuRb\
sample with $74.9\unit{mass-\%}$ \fuRb\ and $25.1\unit{mass-\%}$
OsO$_2$ and an OsO$_2$ sample. It illustrates the significant
additional heat capacity present in \fuRb. The difference between
$C$ and the shaded region is $74.9\unit{\%}$ of the intrinsic heat
capacity of \fuRb.}
  \end{center}
\end{figure}

The full temperature dependence of the intrinsic heat capacity of
\fuRb\ is therefore obtained by subtracting the appropriate amount
of the OsO$_2$ heat capacity contribution according to $C_1 =
\eta_m^{-1} C - (\eta_m^{-1}-1) C_2$. Here, $C_1$ is the heat
capacity of \fuRb, $C_2$ of OsO$_2$, and $C$ is the measured heat
capacity of the mixed system, all in energy per temperature per
mass. To this end we have measured a sample of the starting
material OsO$_2$. Its heat capacity is shown in
Fig.~\ref{fig:CdivT3_vs_logT_Rb_and_OsO2} ($C_2$) together with
the heat capacity of a \fuRb\ sample ($C$) with a superconducting
mass fraction of $\eta_m = 74.9\unit{\%}$. Hence, the difference
between the mixed system heat capacity $C$ and $25.1\unit{\%}$ of
$C_2$ (shaded region) is $74.9\unit{\%}$ of the intrinsic heat
capacity of \fuRb\ $C_1$. The Sommerfeld coefficients used in the
plot are $\sfc = 27.5 \unit{\mu J/(g\,K^2)}$ for OsO$_2$ and $\sfc
= 70.3 \unit{\mu J/(g\,K^2)}$ for the \fuRb\ sample.

\begin{figure}
  \begin{center}
  \includegraphics{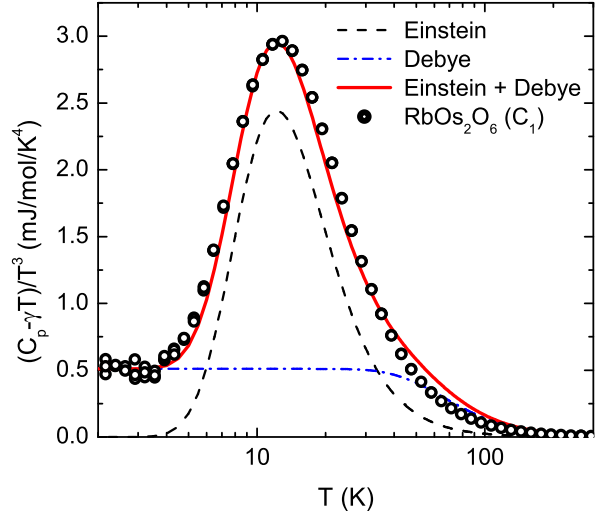}
  \caption{\label{fig:spectrumRb} (Color online) Decomposition of the lattice heat
capacity $(\Cp-\sfc T)/T^3$ of \fuRb\ into Debye and Einstein-mode
components. In such a plot, an Einstein contribution to the heat
capacity appears as a bell-shaped feature with a maximum at $T =
T_E/4.93$.\cite{Ch1961} The data are well described by a combined
Debye-Einstein model with $\Theta_\mathrm{D}=325\unit{K}$,
$T_\mathrm{E}=60\unit{K}$, and $n\approx 0.33\cdot 9$ modes per
f.u.}
  \end{center}
\end{figure}

The resulting specific heat of \fuRb\ ($C_1$) on a logarithmic
temperature scale is shown in Fig.~\ref{fig:spectrumRb}. In such a
plot, an Einstein contribution to the heat capacity appears as a
bell-shaped feature with a peak at $T = T_E/4.93$.\cite{Ch1961}
The data clearly indicate such a contribution with an Einstein
temperature of $60\unit{K}$ and a density of $0.33 \cdot 9
\unit{modes/f.u.}$. Assuming this phonon involves the displacement
of the Rb atoms, this means an effective number of $3$ modes per
Rb ion. Three modes are compatible with a 3-dimensional potential
for the Rb ions to move in as is expected from the tetrahedral
symmetry of the Rb site.

\begin{figure}
  \begin{center}
  \includegraphics{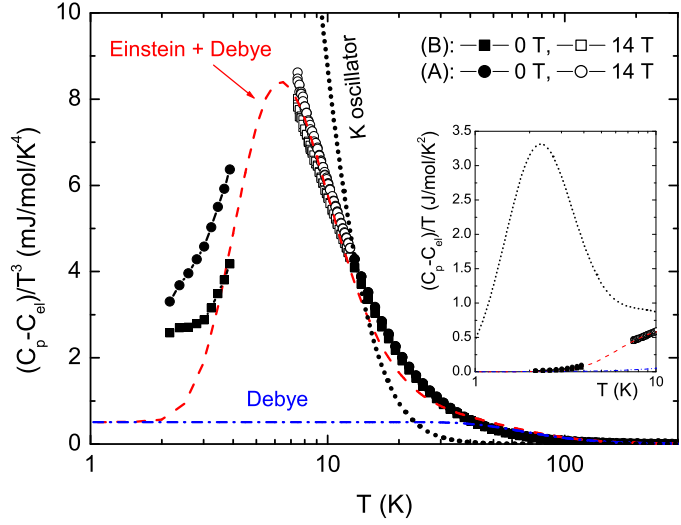}
  \caption{\label{fig:spectrumK} (Color online) $(\Cp-C_\mathrm{el})/T^3$ of \fuK\
on a logarithmic temperature scale. The dashed line shows a
combined Einstein-Debye contribution with $n=0.15 \cdot
9\unit{modes/f.u.}$, $T_\mathrm{E}=31\unit{K}$, and
$\Theta_\mathrm{D}=325\unit{K}$ to illustrate a best fit to the
data if an Einstein-type contribution is assumed. We use the Debye
temperature from \fuRb\ (Fig.~\ref{fig:spectrumRb}). The dotted
line indicates the large lattice contribution expected at low
temperatures from a K ion in an anharmonic
potential.\cite{KuPi2006} The absence of such a contribution may
indicate a freezing of the rattling motion at $T_{\mathrm{p},i}$.
(A) and (B) refer to the two extrapolations used to extract
$\gamma$ as described in the main text. }
  \end{center}
\end{figure}

As the \fuK\ samples are high-quality single crystals, the \fuK\
data can be analyzed as measured. We have observed a vanishingly
small residual Sommerfeld coefficient ($\Cp/T \to -0.35\pm0.4
\unit{mJ/(mol\,K^2)}$ as $H \to 0\unit{Oe}$ at $0.46\unit{K}$) as
expected from a single phase sample and indicative of a fully
gapped electronic excitation spectrum. The temperature dependent
lattice heat capacity data is shown in Fig.~\ref{fig:spectrumK}.
In the normal state above $6.2\unit{K}$, $\sfc T$ was used for the
electronic heat capacity $C_\mathrm{el}$. The low temperature data
from $2$ to $3.8\unit{K}$ were obtained using the electronic heat
capacity of an isotropic superconductor
$C_\mathrm{el}=C_\mathrm{es}=8.5 \, \gamma \, T_\mathrm{c}
\exp(-1.44 \cdot 2\Delta(0\unit{K})/3.53 \cdot T_\mathrm{c}/T)$.
The data does not fit as nicely to a combined Debye-Einstein model
as does the data for \fuRb. This might be due to the alkali-ion
potential being very anharmonic, leading to a vibrational spectrum
that is not simply modeled by an Einstein mode. Nevertheless, we
show a best fit of an Einstein contribution to the data in the
figure, resulting in $n=0.15 \cdot 9\unit{modes/f.u.}$ and
$T_\mathrm{E}=31\unit{K}$. For the Debye contribution which is
expected to result mainly from the rigid Os-O network we have
reused the Debye temperature from \fuRb. The question about the
low-energy lattice dynamics appears to be central to the physics
of \fuK. The calculated anharmonic potential for the K ion results
in a set of discrete vibrational energies, leading to a distinct
$T$ dependence of the specific heat.\cite{KuPi2006} The data
deviates markedly from this model calculation at low temperature
(Fig.~\ref{fig:spectrumK}). Possibly, this indicates a freezing of
the dynamic motion of the alkali ion.

\subsection{Resistivity}

\begin{figure}
  \begin{center}
  \includegraphics[width=0.6\textwidth]{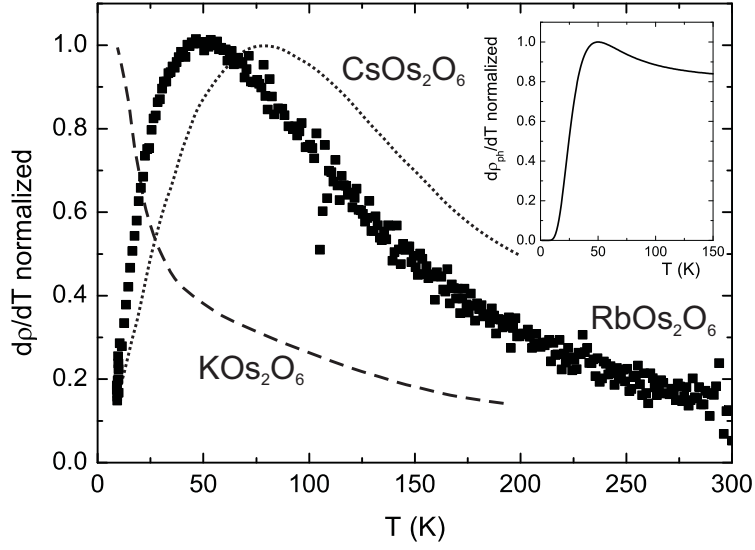}
  \caption{\label{fig:drhodT_for_pyrochlore_SC} Temperature
derivative of the electrical resistivity
$\mathrm{d}\rho/\mathrm{d}T$ normalized to the value at its
maximum showing the systematic variation of the peak location with
the $A$ ion. The data for \fuCs\ (dotted curve) and \fuK\ (dashed
curve) are taken from Ref.~\cite{YoMuHi2004} and
Ref.~\cite{YoMuMaHi2004} respectively. The two curves are
smoothed. The inset shows the calculation for a single Einstein
phonon.}
  \end{center}
\end{figure}

As has been noted before,\cite{YoMuMaHi2004} the resistivity of
\fuK\ shows a peculiar downward curvature, which extends to the
lowest measured temperatures. This downward curvature also exists
in the other \fua\ compounds, though at different temperatures:
the curve looks like an additional hump superimposed on a more
smooth background, the hump having its peak at a temperature
shifting systematically to higher temperatures on going from $A$=K
to $A$=Cs. We analyze this behavior by taking the derivative of
the resistivity with respect to temperature and then locating the
maximum of the resulting curve. Owing to varying effective
geometric factors due to grain sizes, the absolute values of the
resistivities among samples vary significantly. In
Fig.~\ref{fig:drhodT_for_pyrochlore_SC} we therefore show the
normalized derivative of the resistivity where we have set the
peak value to unity.

For illustrative purposes, we compare these results with a model
calculation for the resistivity caused by an Einstein
phonon:\cite{Engquist1980}
\begin{equation}\label{eqn:rhoofEinsteinphonon}
  \rho_\mathrm{ph} = \frac{2\pi m^*\alpha^2F_\mathrm{E}}{ne^2} \frac{\coth\left(T_\mathrm{E}/(2T)\right)}{1+2/3 \sinh^2\left(T_\mathrm{E}/(2T)\right)},
\end{equation}
where $m^*$ is the averaged band mass, $\alpha^2 F_\mathrm{E}$ is
an Einstein spectral function, $n$ is the number of conduction
electrons per unit volume, $e$ is the elementary charge, and
$k_\mathrm{B} T_\mathrm{E}$ is the energy of the Einstein mode.
The inset of Fig.~\ref{fig:drhodT_for_pyrochlore_SC} shows the
derivative of the model resistivity for a peak at $50\unit{K}$. It
describes the data well at low temperatures, but deviates at
higher temperatures, where the model predicts a resistivity linear
in $T$.

\begin{figure}
  \begin{center}
  \includegraphics{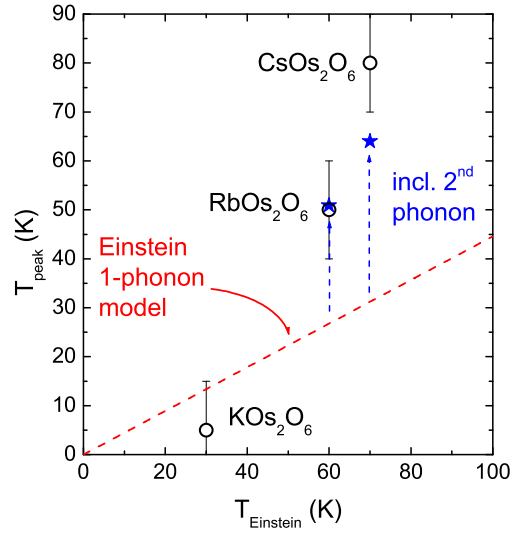}
  \caption{\label{fig:drhodT_for_pyrochlore_SC_PD} (Color online) Systematic
variation with the $A$ ion of the maximum in the temperature
derivative of the resistivity $T_\mathrm{peak}$ versus the
characteristic temperature of the low-energy phonon
$T_\mathrm{Einstein}$ extracted from heat capacity measurements.
For illustration we show where the maxima are expected to lie
according to the solution of the linearized Boltzmann equation for
a single phonon by a dashed line.\cite{Engquist1980} Inclusion of
a second phonon by Matthiessen's rule at around $140\unit{K}$ for
\fuRb\ and around $175\unit{K}$ for \fuCs
\cite{HiYoMuYaMu2005,HiYoYaMuMaMu2005} moves the peak temperature
up towards the measured value shown by the blue stars and arrows.}
  \end{center}
\end{figure}

The resulting temperatures $T_\mathrm{peak}$ at which
$\mathrm{d}\rho/\mathrm{d}T$ peaks are plotted in
Fig.~\ref{fig:drhodT_for_pyrochlore_SC_PD} versus
$T_\mathrm{Einstein}$, the energy of the Einstein phonon mode
extracted from the heat capacity measurement. $T_\mathrm{Einstein}
\approx 70\unit{K}$ for \fuCs\ is taken from
Ref.~\cite{HiYoMuYaMu2005,HiYoYaMuMaMu2005}. It can be shown that
the derivative of the model resistivity
$\mathrm{d}\rho_\mathrm{ph}/\mathrm{d}T$ shows a maximum at
$T_\mathrm{peak} = T_\mathrm{E}/a$, where $a \approx
2.243$.\footnote{$a$ is the nonzero solution of $2\sinh(a/2) +
5a\cosh(a/2) - 18a\cosh^3(a/2) + 16\sinh(a/2)\cosh^6(a/2) -
8a\cosh^7(a/2) + 12a\cosh^5(a/2) = 0$.} For illustrative purposes,
this line is shown along the experimental results from \fua\ in
Fig.~\ref{fig:drhodT_for_pyrochlore_SC_PD}. As one might expect,
this simple Einstein phonon model is not sufficient to describe
the entire $\rho(T)$ curves, because scattering at other phonons
is not included. If a second phonon is considered and the
resulting resistivity is assumed to be simply the sum of the two
single-phonon results according to Matthiessen's rule, then the
location of the peak moves more toward the measured points: in an
analysis of the specific heat data, Hiroi \textit{et
al}.\cite{HiYoMuYaMu2005,HiYoYaMuMaMu2005} have explained their
data using a second phonon at around $140\unit{K}$ for \fuRb\ and
around $175\unit{K}$ for \fuCs. Including a phonon at these
temperatures moves the peak location up in temperature to
$51\unit{K}$ for \fuRb\ and $64\unit{K}$ for \fuCs, shown by blue
stars and arrows in Fig.~\ref{fig:drhodT_for_pyrochlore_SC_PD}.
The deviation of the \fuCs\ $T_\mathrm{peak}$ towards even higher
temperatures may be due to the fact that at these temperatures a
significant contribution to the phonon spectrum already stems from
the Debye-phonons neglected up to now in the transport model. The
overall trend, however, is well represented by such an analysis
and it points to the significant scattering by the $A$ atoms.

\section{Discussion}



The systematic variation of the position of the maximum slope in
the resistivity indicates a close connection between the dynamics
of the alkali ion $A$ and the electronic transport properties.
\fuK\ is somewhat different from the other two compounds, because
its peak temperature is lower than expected from the simple
Einstein phonon model
(Fig.~\ref{fig:drhodT_for_pyrochlore_SC_PD}). This might be a
result of the strong anharmonicity of the cage potential. It is
known that the alkali ions move in a strongly anharmonic potential
and couple to the conduction electrons owing to their large
excursion from equilibrium, and previous results have been
interpreted in this
scenario.\cite{HiYoMuYaMu2005,HiYoYaMuMaMu2005}. This results in
the electron-phonon coupling parameter $\lambda_\mathrm{ep}$ to
increase, driving \fuK\ more towards the strong-coupling regime.
Electron scattering from the alkali mode as the reason for the
downward curvature of the resistivity has been suggested before by
Kune$\check{\mathrm{s}}$ \textit{et al}.,
(Ref.~\cite{KuJePi2004}). While the scattering at phonons seems to
be the most plausible mechanism at this point, other scenarios for
the peculiar behavior of the resistivity are also conceivable:
Fermi surface nesting has been found in \fuK\ and has been
proposed as the driving force for strong spin
fluctuations.\cite{SaMeYeShFr2004} It remains to be worked out in
detail to what degree such fluctuations or even more exotic
excitations could account for the measured resistivity. We note,
however, that the static susceptibility of \fuK\ is not
significantly enhanced; rather it is smaller than expected based
on the band density of states.


In the following we try to identify the various contributions to
the measured mass enhancement reflected in the Sommerfeld
coefficient $\sfc$. It is known that Coulomb and electron-phonon
effects in a metal can be combined in a multiplicative
fashion.\cite{PrSa1967} An interpretation of the additional
enhancement in terms of Coulomb correlations would therefore mean
a parametrization according to $1 + \lambda = (1 +
\lambda_\mathrm{ep})(1 + \lambda_\mathrm{c})$. In this
interpretation the over-all bandwidth is reduced by a factor of
$1+\lambda_\mathrm{c}$ due to electron-electron interactions. The
resulting Coulomb enhancement parameter $\lambda_\mathrm{c}$ for
\fuRb\ is $\approx 1.0$ and for \fuK\ it is $\approx 1.7$
\{$4.3$\}, from which we estimate the interaction strength:
assuming a conduction electron density of $n = 2$ electrons per
Os, we can make a crude estimate of the Coulomb interaction
potential: $V_\mathrm{c} = n \lambda_\mathrm{c}/N(0) \approx 2
\unit{eV}$ for \fuRb\ and $3\unit{eV}$ \{$9\unit{eV}$\} for \fuK.
Here, $N(0) = g(E_\mathrm{F})/2 $ is the density of states at the
Fermi level for one spin direction. Both of these values indicate
a large electron correlation.


It is instructive to compare the \fua\ series with a Ru based
$\alpha$-pyrochlore series, where variations of the Ru-O-Ru angle
results in a drastic change of the density of states. The compound
Y$_{2-x}$Bi$_x$Ru$_2$O$_7$ is a Mott insulator for small $x$ and a
normal paramagnetic metal for $x\approx2$. The band Sommerfeld
coefficient of Bi$_2$Ru$_2$O$_7$ is calculated to be
$8.0\unit{mJ/(mol\,K^2)}$\cite{IsOg2000}, resulting in a very
small mass enhancement ($\lambda \approx 0.23$). The correlations
increase upon Y substitution resulting in an increase of the
Sommerfeld coefficient from about $10\unit{mJ/(mol\,K^2)}$ at $x
\approx 2$ to almost $90\unit{mJ/(mol\,K^2)}$ at $x \approx
0.9$.\cite{YoSa1999} The enhancement at $x=0.9$ is thus
substantial: $\lambda \approx 10$. Below this critical value
correlations become so strong that a gap opens in the density of
states and the Sommerfeld coefficient decreases rapidly and
vanishes in the Mott insulating state. Using the same arguments as
above for \fua, the interaction potential $V_\mathrm{c}$ in
Y$_{1.1}$Bi$_{0.9}$Ru$_2$O$_7$ is of the order of $10 \unit{eV}$,
certainly large enough to cause a gap for a bandwidth of typically
$3\unit{eV}$. At the same time the Ru-O-Ru angle is reduced from
about $139\unit{^{\circ}}$ for Bi$_2$Ru$_2$O$_7$ to
$129\unit{^{\circ}}$ for Y$_2$Ru$_2$O$_7$.\cite{KaTaYaKaYa1993}
This strongly indicates that a decreasing $B$-O-$B$ angle causes
the band to narrow and the correlations among conduction electrons
to increase. \fua\ shows the same tendency: the Os-O-Os angle in
\fuRb\ is $139.4\unit{^{\circ}}$ [$\sfc=44\unit{mJ/(mol\,K^2)}$],
while that for \fuK\ is $137.9\unit{^{\circ}}$ [$\sfc=76$ to $110
\unit{mJ/(mol\,K^2)}$]. We thus interpret the mass of \fua\ to be
increasingly enhanced by electron correlations going from $A$=Cs
to $A$=K. Possibly the correlations are especially strong in some
subband, leading to a CDW like or orbitally ordered state at
around $7\unit{K}$ ($T_\mathrm{p,1}$ and $T_\mathrm{p,2}$ in
Fig.~\ref{fig:DeltaCatTc}).

This is consistent with another series of pyrochlores: according
to Solovyev (Ref.~\cite{solovyev2003}), a decreasing Mo-O-Mo angle
in $R_2$Mo$_2$O$_7$ reduces the interaction between
Mo(t$_\mathrm{2g}$) orbitals which is mediated by O(2p) states,
causing diminishing overlap and thus enhanced mass. Although this
author finds the lattice parameter to be the relevant parameter
determining the physics of the compound, the physical properties
of various pyrochlores appear to be more appropriately
parameterized by the Mo-O-Mo angle.



An effective Hubbard $U$ of $2\unit{eV}$ is in line with other
transition metals. According to \cite{SiBlSchSo2002}, this is not
enough to classify Cd$_2$Os$_2$O$_7$ as a strongly correlated
system. In Ref.~\cite{solovyev2003}, however, a $U$ of $1.5$ to
$2.5\unit{eV}$ explains the Mott-insulating properties of
$R_2$Mo$_2$O$_7$ very well. This might be a somewhat weak
argument, because Cd$_2$Os$_2$O$_7$ is a 5d material while
$R_2$Mo$_2$O$_7$ is 4d, which are usually more strongly
correlated. On the other hand, the LDA bandwidth of \fua\ is in
the region of $W \approx
3\unit{eV}$\cite{KuJePi2004,SaMeYeShFr2004} and the interaction
potential $V_\mathrm{c}$ is therefore comparable to $W$ and thus
the Mott-Hubbard localization concept is applicable. The anomalous
pressure dependence of the superconducting transition temperature
is in line with these findings:\cite{MuTaTeTaToYoMuHi2005}
Hydrostatic pressure reduces the correlations working against
superconductivity, thus initially increasing the critical
temperature. For this reason it would be of interest to see if
NaOs$_2$O$_6$ shows metallic or insulating behavior.


It is enlightening to compare the electronic structure of \fua\ to
the one of OsO$_2$, since, in a (limited) way, OsO$_2$ can be
regarded as \fua\ without the $A$. It has a rutile structure with
the OsO$_6$ octahedra sharing edges in the c direction and corners
in the plane perpendicular to c.\cite{mattheiss2003} While the
Os-Os distance to the neighboring octahedron with the shared edge
and the length of the shared edge itself are shorter than the
Os-Os and O-O distances in \fua, all other Os-O, O-O, and Os-Os
distances in OsO$_2$ are slightly longer than the ones in \fua.
The Os-O-Os angles in OsO$_2$ are $105.0\unit{^{\circ}}$ along the
c axis and $127.5\unit{^{\circ}}$ perpendicular to it. Pyrochlores
with a $B$-O-$B$ angle around $127\unit{^{\circ}}$ usually are in
an insulating state. Directional resistivity measurements on
OsO$_2$ could thus be quite insightful in this respect.

The density of states at the Fermi level for OsO$_2$ of about $13$
to $15\unit{states/(u.c.\,Ry \, spin)}$ from the band structure
calculation in Ref.~\cite{mattheiss2003} results in a Sommerfeld
coefficient $\sfc_\mathrm{b} \approx 2.4\unit{mJ/(mol \, K^2)}$.
We measure a $\sfc$ of about $6.1\unit{mJ/(mol \, K^2)}$,
resulting in a specific heat enhancement of about $1 + \lambda =
6.1/2.4 \approx 2.5$. Assuming a small electron phonon coupling in
OsO$_2$ of $\lambda_\mathrm{ep} \approx 0.2$ and an enhancement of
the form $\sfc/\sfc_\mathrm{b} =
(1+\lambda_\mathrm{ep})(1+\lambda_\mathrm{c})$, this results in an
enhancement due to correlations of $\lambda_\mathrm{c} \approx
1.1$. This is quite a large enhancement parameter similar to the
one of \fua, providing further evidence that the Coulomb
correlations are inherent to the Os-O system.

Recent calculations have shown that the electronic structure does
not change significantly on changing the alkali metal ion
$A$.\cite{SaFr2005} Therefore, all these considerations point to
the need to consider interactions that are not captured in the
electronic structure calculations. In light of the significant
electron-phonon interaction, and even stronger effects due to
electron correlations, it will be of interest to further focus on
the possible role played by the 3D-triangular geometry of the Os-O
network. In view of the strong electron correlations, the
frustrated geometry might be of importance in these materials.

We take the non-superconducting state of
Y$_{2-x}$Bi$_x$Ru$_2$O$_7$ to be a clear indication that the
additional enhancement mechanism is to be regarded separately from
the pairing mechanism. Both \fua\ and Y$_{2-x}$Bi$_x$Ru$_2$O$_7$
show significant enhancement due to correlations, but only \fua\
is superconducting. The question thus remains why \fua\ is
superconducting and Y$_{2-x}$Bi$_x$Ru$_2$O$_7$ is not. The answer
may lie in the electron-phonon coupling, which is unusually large
in \fua. We expect this to be due the $\beta$-pyrochlore instead
of the $\alpha$-pyrochlore structure, which leaves the 16d site
empty and the 8b site occupied by $A$ instead of an oxygen atom.
This should significantly modify the phonon spectrum, resulting in
a large $\lambda_\mathrm{ep}$.

\section{Conclusion}

Our data from high quality single crystalline \fuK\ show that
\fuK\ is a particularly interesting transition metal oxide: it is
an intermediate to strong coupling type-II superconductor with a
coupling parameter $\lambda_\mathrm{ep} \approx 1$ to $1.6$. \fuK\
has a high Sommerfeld coefficient for a pyrochlore of $76$ to $110
\unit{mJ/(mol\,K^2)}$. We estimate a Ginzburg-Landau coherence
length $\xi \approx 31$ to $37\unit{\AA}$, about half the one in
\fuRb\ due to the renormalization of the Fermi velocity
$v_\mathrm{F}$. We estimate the condensation energy $\ce = 2.0$ to
$2.5 \unit{J/mol}$. The effective mass, even after the measured
strong electron-phonon renormalization is taken into account, is
threefold enhanced over the LDA band mass. We interpret this as
due to Coulomb correlations. The renormalization affects those
electrons that are paired to form the superconducting condensate.
While for \fuRb\ the additional heat capacity can be well
characterized by an Einstein model, the contribution for \fuK\ is
somewhat more unusual, possibly due to strong anharmonicity. We
associate this special phonon with a rattling motion of the alkali
ions, resulting in three modes per Rb for \fuRb. In \fuK, the
absence of lattice heat capacity at low temperatures may indicate
the freezing of this motion. The dynamics of the alkali ions
causes large scattering of the charge carriers which shows up in
an unusual temperature dependence of the electrical resistivity
which varies systematically with the alkali ion.

\section{Acknowledgment}

We thank J.~Kune$\check{\mathrm{s}}$ and W.~E.~Pickett for helpful
discussions on the calculation of the lattice vibrational spectrum
and R.~Saniz for providing revised values for the Fermi
velocities. This study was partly supported by the Swiss National
Science Foundation.


\newpage


\newpage

\begin{figure}
  \begin{center}
  \caption{\label{fig:CvsHandCrystals} (Color online) Heat capacity versus magnetic
field of \fuK\ at $0.46\unit{K}$. An extrapolation to the upper
critical field $H_\mathrm{c2} \approx 24\unit{T}$ \{$35\unit{T}$\}
results in a Sommerfeld coefficient $\sfc=76\unit{mJ/(mol\,K^2)}$
\{$\sfc=110\unit{mJ/(mol\,K^2)}$\}. The inset shows the measured
crystals.}
  \end{center}
\end{figure}

\newpage

\newpage

\newpage

\newpage

\newpage

\newpage

\newpage

\newpage

\end{document}